\documentclass[a4paper]{AO4ELT}  

\usepackage{microtype}
\usepackage{biblatex}
\usepackage{amsmath,amsfonts,amssymb}
\usepackage{graphicx}
\usepackage{pst-all} 
\usepackage[colorlinks=true, allcolors=blue]{hyperref}
\usepackage[hang]{subfigure}
\usepackage{multirow}
\usepackage{aasmacros}

\makeatletter         
\def\@maketitle{
\includegraphics[width = 165mm]{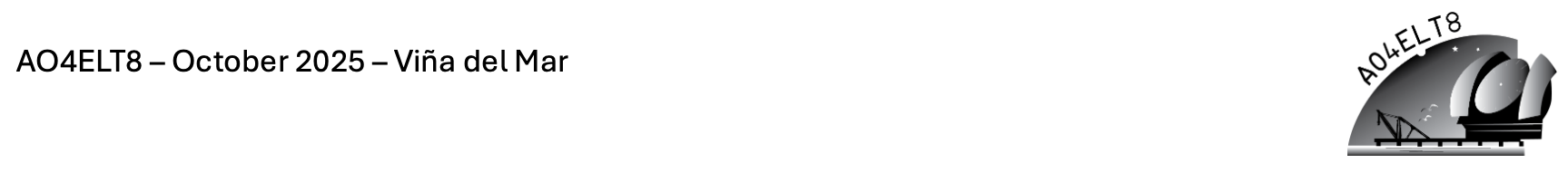}\\[8ex]
\begin{center}
{\Huge \bfseries \sffamily \@title }\\[4ex] 
{\large  \@author}\\[4ex] 
\@date
\end{center}}
\makeatother

\title{MORFEO: Advancing Towards Final Design}

\author[a]{Lorenzo Busoni}
\author[a]{Guido Agapito}
\author[a]{Marco Bonaglia}
\author[a]{Alfio Puglisi}
\author[a]{Marco Xompero}
\author[a]{Matteo Aliverti}
\author[a]{Francesca Annibali}
\author[a]{Carmelo Arcidiacono}
\author[a]{Natalia Auricchio}
\author[a]{Nicolò Azzaroli}
% ...

\affil[a]{Istituto Nazionale di Astrofisica (INAF), Italy}
\affil[b]{Herzberg Astronomy and Astrophysics Research Centre, National Research Council of Canada, Victoria, BC, Canada}
\affil[c]{Institut de Planétologie et d'Astrophysique de Grenoble (IPAG), CNRS, Univ. Grenoble Alpes, Grenoble, France}
\affil[d]{University of Galway, Galway, Ireland}
\affil[e]{National Astronomical Observatory of Japan (NAOJ), Tokyo, Japan}
\affil[f]{University of Tokyo, Tokyo, Japan}
\affil[g]{Kyoto University, Kyoto, Japan}
\affil[h]{Institut de Physique des 2 Infinis de Lyon (IP2I), Univ. Lyon, Lyon, France}
\affil[i]{Osaka University, Osaka, Japan}
\affil[j]{European Southern Observatory (ESO), Garching, Germany}

\authorinfo{Further author information: Lorenzo Busoni: E-mail: lorenzo.busoni@inaf.it}

\begin{document} 
\maketitle

\begin{abstract}
The Multiconjugate adaptive Optics Relay For ELT Observations (MORFEO) is a first-generation adaptive optics module for the Extremely Large Telescope (ELT), designed to deliver a diffraction-limited, highly uniform 53x53 arcsec field of view to the MICADO near-infrared camera. As the project advances toward its Final Design Review (FDR), significant consolidations have been achieved across all subsystems. This paper presents an updated overview of the MORFEO system, highlighting its dual operational modes (MCAO and SCAO) and recent developments in its opto-mechanical architecture. We dedicate specific focus to the core adaptive hardware, detailing the fifth-generation post-focal deformable mirrors, the highly complex Laser Guide Star (LGS) objective zoom system required to track sodium layer variations, and the Natural Guide Star (NGS) low-order and reference sensing strategies. Furthermore, we detail the advanced pseudo-open-loop control strategy managed by a split Hard and Soft Real-Time Computer architecture. Finally, we report the latest end-to-end performance estimations obtained via the SPECULA simulation framework, demonstrating compliance with the stringent Strehl Ratio and sky coverage requirements under median atmospheric conditions.
\end{abstract}

\keywords{Adaptive Optics, ELT, MORFEO, MCAO, LGS, NGS, Wavefront Control, Deformable Mirrors}

\section{INTRODUCTION}
\label{sec:intro}

The Multiconjugate adaptive Optics Relay For ELT Observations (MORFEO, formerly known as MAORY) \cite{2021Msngr.182...13C} is the adaptive optics module designed to provide a wide, uniformly corrected field of view for the Extremely Large Telescope (ELT) \cite{2024Msngr.192....3C}. Its primary objective is to feed the first-light instrument MICADO (Multi-AO Imaging Camera for Deep Observations) \cite{2021Msngr.182...17D} with high-quality, diffraction-limited images in the near-infrared.

Since the preliminary design phase, the MORFEO project has successfully progressed through critical design stages and is currently advancing towards its Final Design Review (FDR) \cite{2024SPIE13097E..22C, BusoniAO4ELT8}. This paper outlines the recent system-level updates, focusing on the core adaptive hardware (wavefront sensors and deformable mirrors), the opto-mechanical evolution, the refined Adaptive Optics (AO) control strategy, and the advanced numerical simulations used to consolidate the error budget.

\section{SYSTEM OVERVIEW AND OPTO-MECHANICAL DESIGN}
\label{sec:system_overview}
MORFEO is a massive and highly complex instrument, with dimensions of approximately $6 \times 6 \times 5$ meters and a total mass of up to 30 tons, situated on the ELT Nasmyth B platform \cite{2024SPIE13097E..22C}. 

The mechanical architecture relies on two fundamental structural groups: the Main Support Structure (MSS), which acts as the primary interface and is kinematically mounted onto the ELT Nasmyth platform to remain unaffected by external deformations, and the Optomechanical Support Structures (OSS), which are hosted on the MSS and hold the sensitive optical components \cite{2024SPIE13096E..5LD}.

\subsection{Relay Optics and Thermal Control}
The Relay Optics design provides a highly stable optical path, operating within a passive thermal cover to ensure thermal homogeneity with the environment, complemented by an active liquid cooling system to extract heat from the main dissipative units \cite{2024SPIE13097E..54M, 2024SPIE13097E..4RP, 2022SPIE12185E..4PA}. 

At the hardware level, the Instrument Control Hardware (ICH) relies on Beckhoff Programmable Logic Controllers (PLCs) distributed via an EtherCAT fieldbus. The ICH independently drives approximately 50 motorized axes and manages interlock safety through dedicated Safety PLCs (using TwinSAFE technology) and hardware modules \cite{2024SPIE13096E..5LD}.

\begin{figure}[ht]
   \begin{center}
   \includegraphics[width=0.8\textwidth]{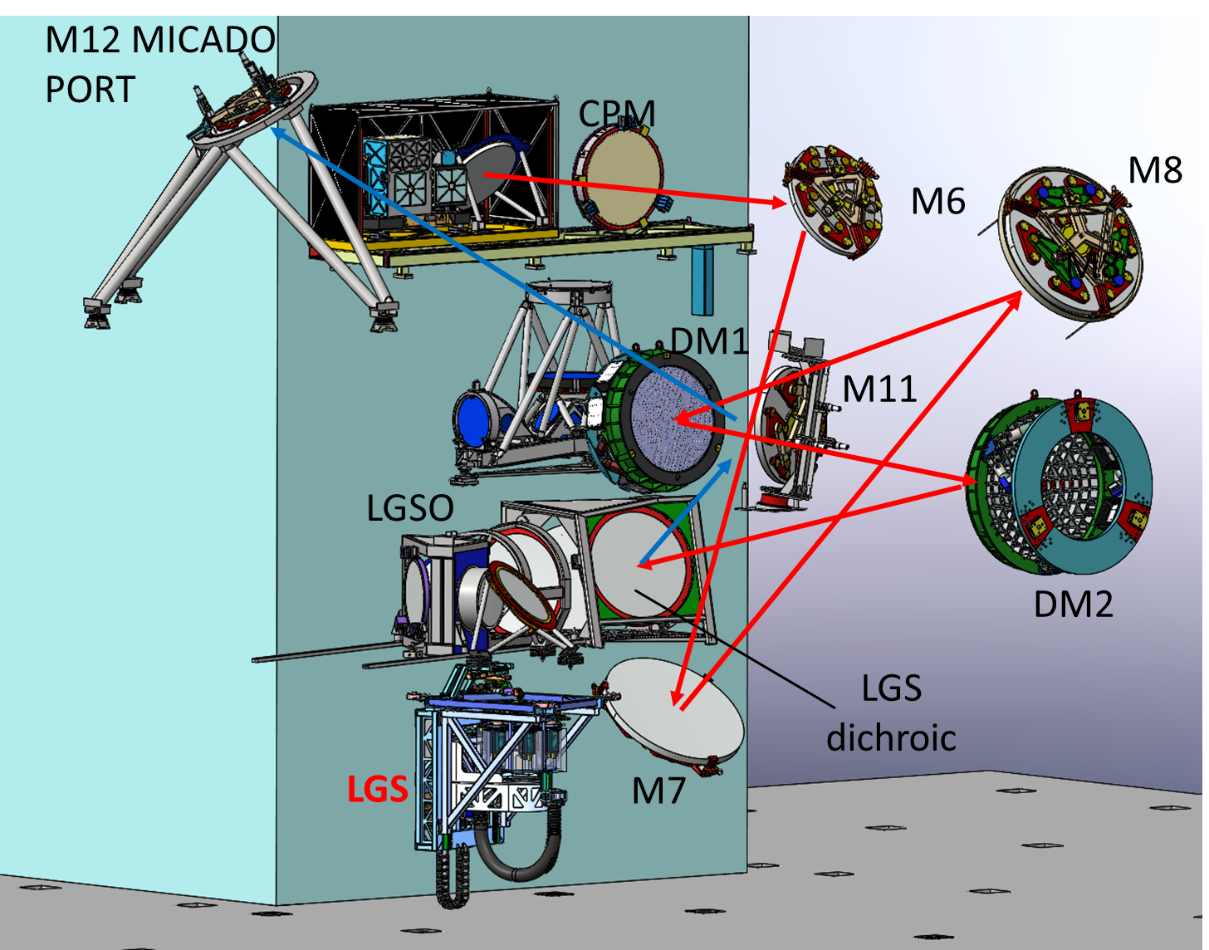}
   \end{center}
   \caption{\label{fig:layout} Updated optical layout and main structure overview of the MORFEO system \cite{2024SPIE13096E..5LD}. The figure highlights the decoupled MSS and OSS, along with the LGS and NGS optical paths.}
\end{figure}

\section{DEFORMABLE MIRRORS AND WAVEFRONT SENSORS}
\label{sec:sensors_dm}
To achieve the stringent MCAO performance requirements, MORFEO relies on a suite of wavefront correction and sensing hardware, namely 3 deformable mirrors conjugated to different atmospheric layers and 12 WFSs probing the turbulent volume by means of 6 LGS and 3 NGS.

\subsection{Post-Focal Deformable Mirrors}
While the ELT's M4 adaptive mirror provides the ground-layer correction, MORFEO incorporates two post-focal deformable mirrors (DM1 and DM2) embedded within the relay to correct for high-altitude turbulence \cite{2021Msngr.182...13C}. These DMs are based on the fifth-generation voice-coil technology developed by AdOptica \cite{BiasiAO4ELT8}.

\begin{figure}[ht]
    \begin{center}
        \subfigure[DM1 full view rendering.
        \label{fig:dm1_01}]
        {\includegraphics[width=0.4\columnwidth]{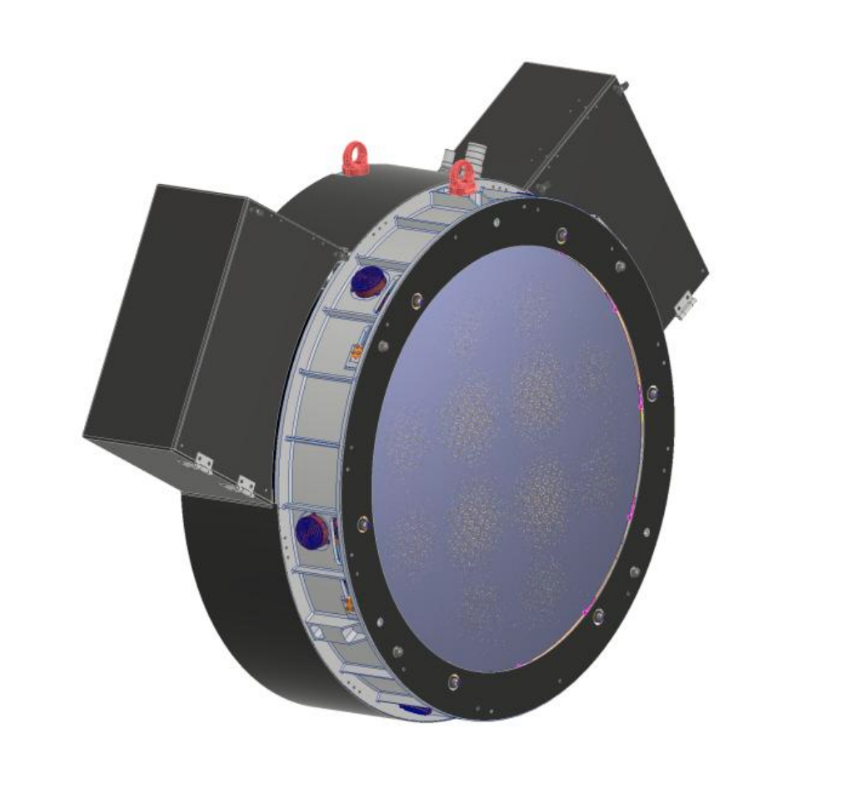}}
        \subfigure[DM1 half view rendering.
        \label{fig:dm1_02}]
        {\includegraphics[width=0.4\columnwidth]{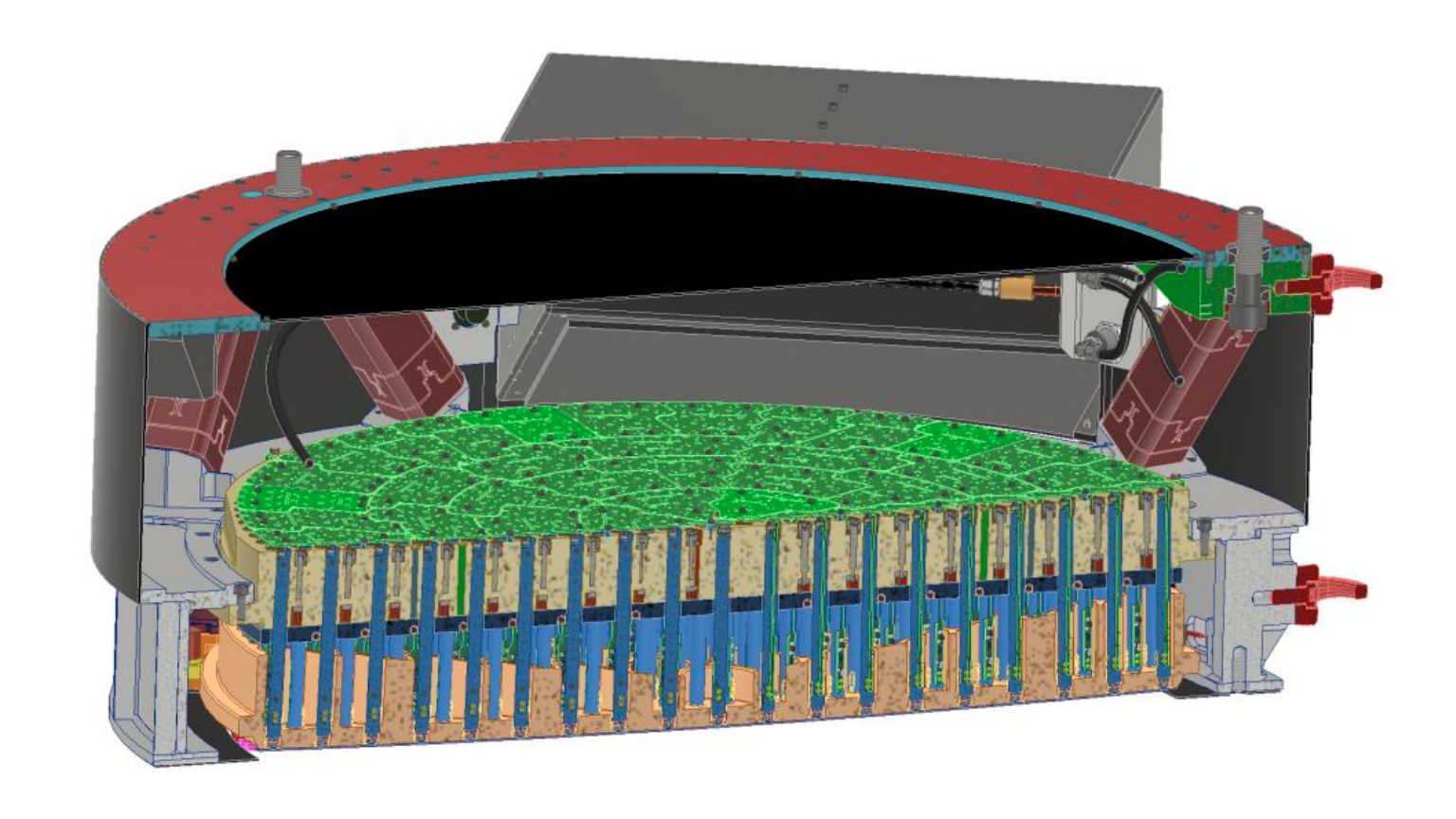}}
    \end{center}
   \caption{\label{fig:dm1} MORFEO DM1.}
\end{figure}
\begin{figure}[ht]
    \begin{center}
        \subfigure[DM2 full view rendering.
        \label{fig:dm2_01}]
        {\includegraphics[width=0.4\columnwidth]{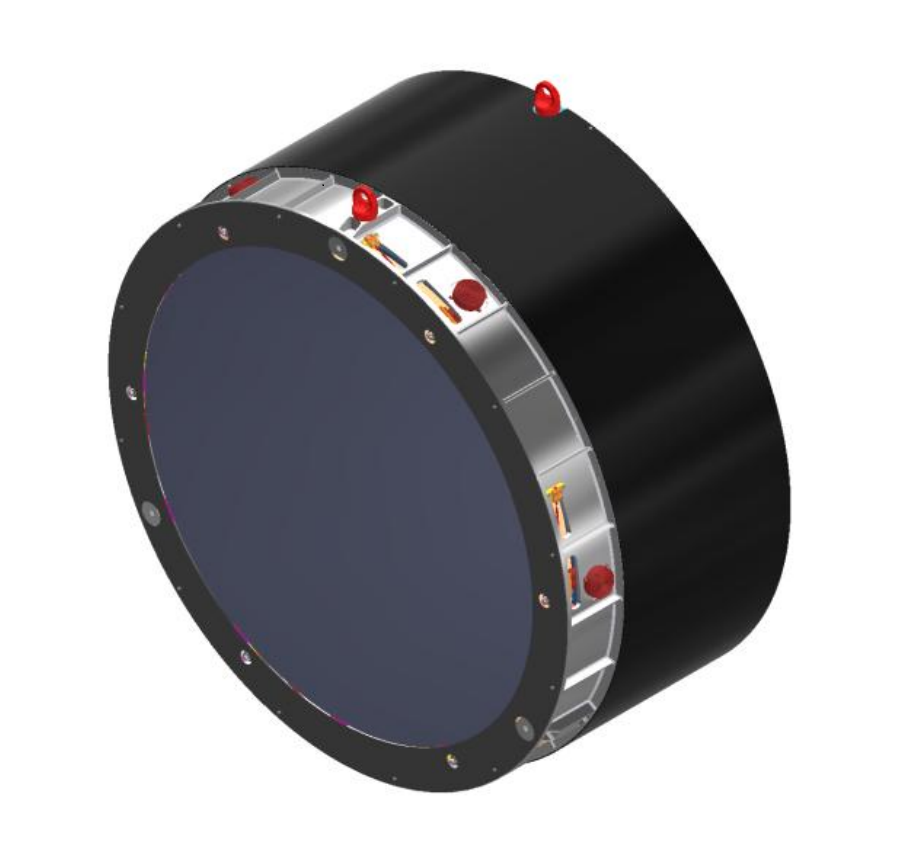}}
        \subfigure[DM2 half view rendering.
        \label{fig:dm2_02}]
        {\includegraphics[width=0.4\columnwidth]{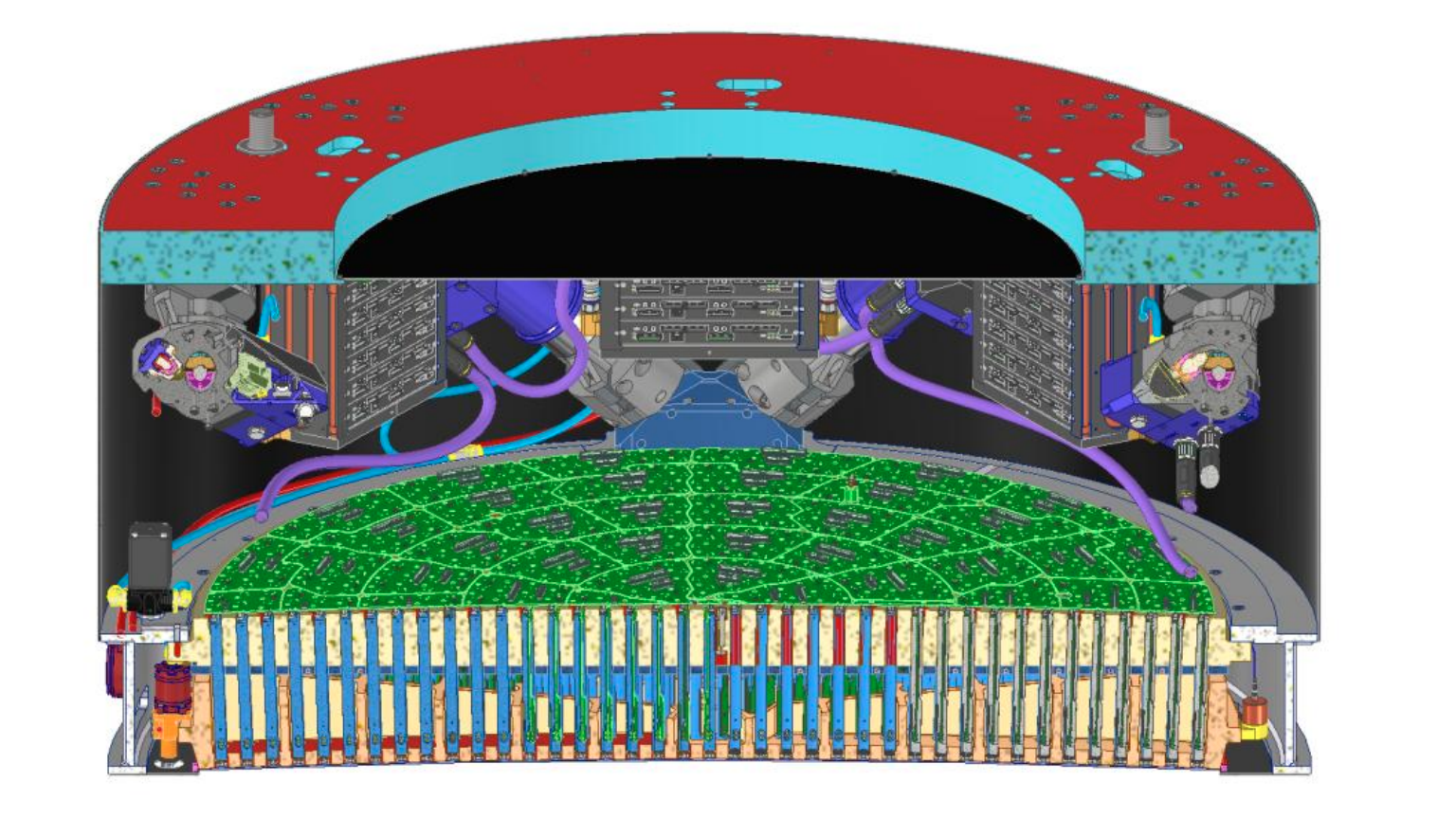}}]
        \end{center}
   \caption{\label{fig:dm2} MORFEO DM2.}
\end{figure}
The DM1 (M9 in the optical path) is a convex 930 mm spherical mirror with 1026 actuators. It is conjugated to about 17500 m altitude with an equivalent pitch of 1.4 m.
The DM2 (M10 in the optical path) is concave, diameter 1224 mm, 1147 actuators (see Fig.~\ref{fig:dm2}). It is conjugated to about 6500 m and the equivalent pitch on the sky is 1.25 m.
Both DMs are intended to provide high performances in terms of speed, quality of correction, stability and stroke. This technology ensures the high actuator density and fast dynamical response times required to operate at the 500 Hz system frame rate, with a fitting error below 40 nm RMS; the DMs will be able to keep a pre-calibrated shape within 10 hour observations time with an error below 25 nm RMS surface. 
The  DMs passed the FDR phase and are now under AIV phase. In particular the DM2 has passed the Integration Readiness Review milestone in December 2025 and it is currently under assembly at AdOptica premises.

\subsection{Laser Guide Star Wavefront Sensors (LGS WFS)}
\label{sec:lgs_wfs}

\begin{figure}[ht]
    \begin{center}
        \subfigure[
        \label{fig:LGS_3D}]
        {\includegraphics[width=0.45\columnwidth]{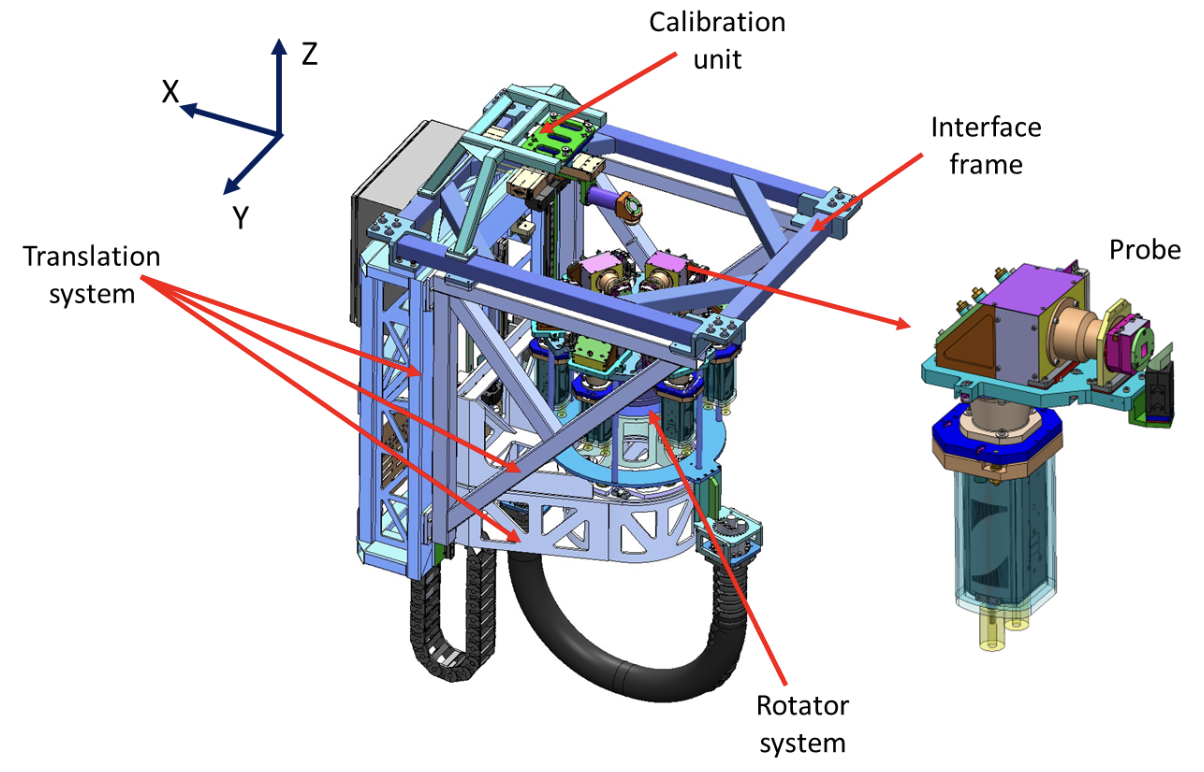}}
        \subfigure[
        \label{fig:LGS_rotation}]
        {\includegraphics[width=0.25\columnwidth]{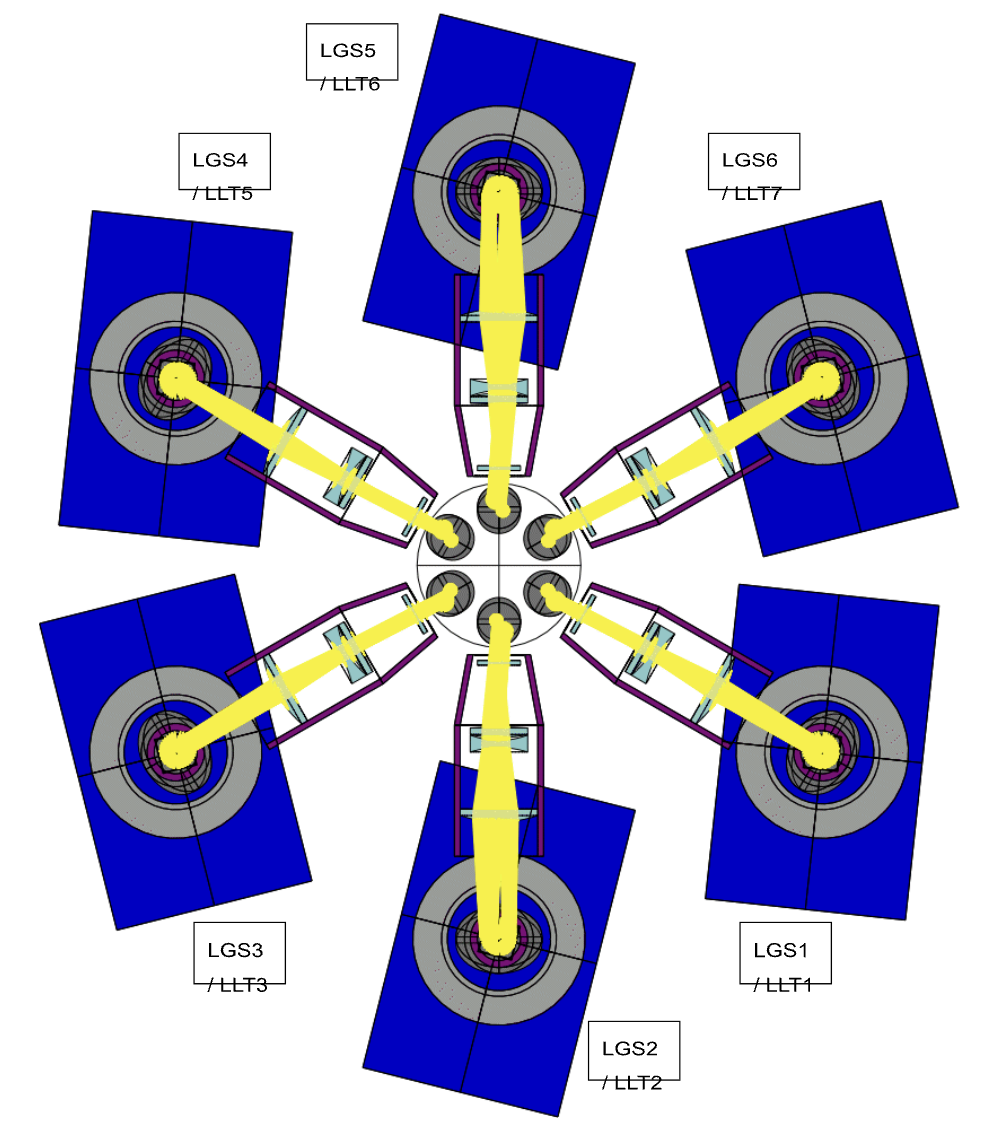}}
        \subfigure[
        \label{fig:LGS_single}]
        {\includegraphics[width=0.21\columnwidth]{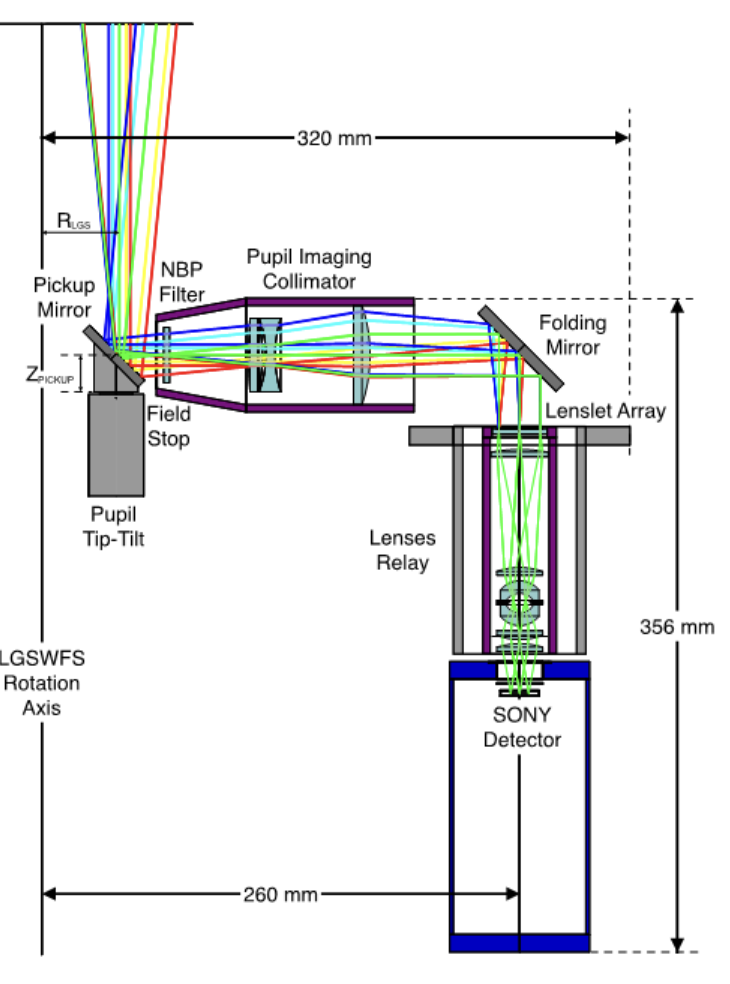}}
        \end{center}
    \caption{Laser Guide Star Wavefront Sensor (LGS-WFS). (a): 3D model of the assembly of six units, with a detailed view of a single unit. (b): top view of the units, showing the beams and the relative rotations. (c): schematic of a single unit, showing the optics and detector.}
    \label{fig:LGSWFS}
\end{figure}

The high-order wavefront sensing is delegated to a constellation of six LGS Shack-Hartmann WFSs arranged on a 45 arcsec radius ring. The entire LGS WFS module is gravity invariant and can be conceptually divided into two optomechanical units: the wavefront sensor probes, which pick off the MORFEO beam after the LGS objective (LGSO) in an LGS focal plane and the support structure that holds the six probes and aligns them within the beam in rotation and focus. The physical arrangement of the support structure and the probes is illustrated in Fig.~\ref{fig:LGSWFS}.

A major opto-mechanical challenge in the LGS path arises from the sodium layer altitude, which varies continuously between roughly 80 km and 200 km due to changes in the telescope Zenith distance and intrinsic mesospheric variability. To compensate for this, the LGSO features a telecentric design to maintain a constant output F number \cite{2024SPIE13096E..5KM} and the LGS support structure is equipped with a linear stage to compensate for the nominal defocus induced by the varying distance, alongside a de-rotator to track telescope elevation and maintain a fixed orientation with respect to the ELT pupil \cite{2018SPIE10703E..1YS}.

Internally, each of the six WFS probes features a pupil steering mirror mounted on a 2-axis piezo stack, allowing precise recentering of the pupil image onto the lenslet array to compensate for possible pupil shifts. The beam then passes through a pupil imaging collimator and a $68 \times 68$ lenslet array, followed by an optical relay that re-images the LGS spots onto the camera detector. The WFS optical design provides a wide Field of View (FoV) of approximately 16 arcsec sampled at $14 \times 14$ pixels per sub-aperture\cite{CostilleAO4ELT7}. This large FoV is critical to minimize spot truncation effects caused by the vertical extension of the sodium layer \cite{2022JATIS...8b1514F}.

Finally, the Sony IMX 425 CMOS detector was selected as the chip for the LGS cameras. A key driver for this choice is its global shutter architecture, which provides a clear performance advantage over rolling shutter sensors for our specific highly dynamic, fast-framerate application \cite{2022JATIS...8b1505A, 2022SPIE12185E..8DA}.


%

\subsection{Natural Guide Star Sensors: Low Order (LO) and Reference (R) WFS}


To resolve tilt anisoplanatism and modes invisible to the LGS constellation (e.g., global tip-tilt, focus, and differential piston from the ELT island effect), MORFEO employs up to three NGSs sensed by three LOR WFS Units \cite{2023aoel.confE..10A}. Each LOR WFS unit splits the incoming NGS light via a VIS/IR dichroic. The Reference (R) WFS operates in the visible to deliver the high-order reference for the LGS tomographic loop, while the LO WFS operates in the near-infrared (J+H band) to penetrate dust and maximize sky coverage \cite{2022JATIS...8b1509P}.

To optimize sensitivity on faint stars, the LO sensing relies on a hybrid baseline architecture: two full-aperture (FA) sensors dedicated to optimal Tip-Tilt tracking, and one $2 \times 2$ Shack-Hartmann sensor specifically configured to isolate and measure focus errors. Furthermore, to address the phase ambiguities across the segmented mirrors (petal modes), the NGS architecture integrates focal-plane sensing methods, such as the Linearized Focal-Plane Technique (LIFT) \cite{PlantetAO4ELT8, 2022SPIE12185E..56A}.

\begin{figure}[ht]
   \begin{center}
   \includegraphics[width=0.95\textwidth]{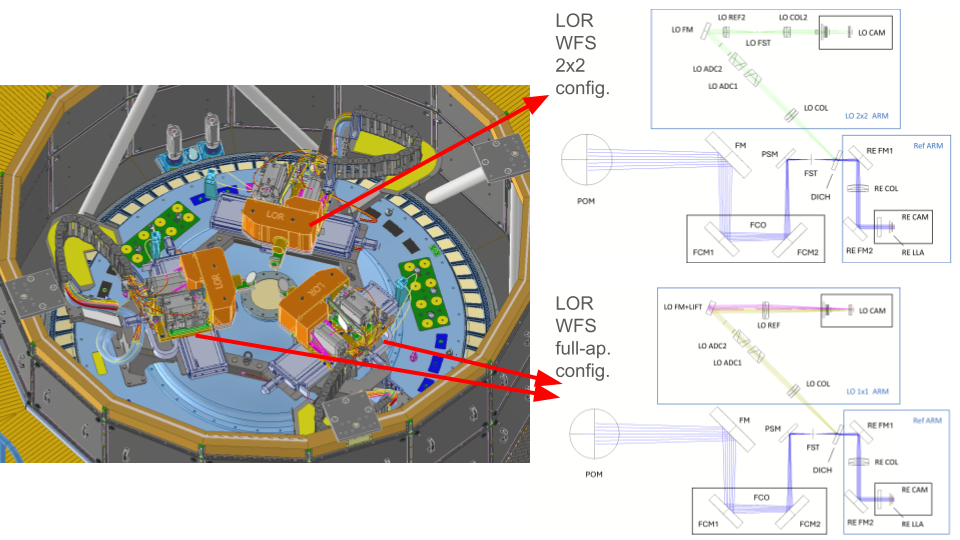}
   \end{center}
   \caption{\label{fig:lor_overview} Left: sketch of the MORFEO Green Doughnut volume with the three LOR WFS Units (orange) on their 2D acquisition stages installed on top of the MICADO cyostat (cyan). Right: optical schematics of the two LOR WFS configurations. Top row: $2\times2$ SH mode; bottom row: full-aperture (FA/LIFT) mode. Both rows show the LO arm lgith path (green: $2\times2$, yellow: full-aperture, pink: LIFT) and REF arm (blue). Key elements: POM – pick-off mirror; FCO – focus compensator; PSM – pupil steering mirror; DICH – dichroic; FM – fold mirrors; LLA – lenslet arrays.}
\end{figure}

\noindent\textbf{Optomechanical design.}
The three LOR WFS Units are mounted within the MORFEO Green Doughnut (GD) on pairs of orthogonal linear acquisition stages arranged in a 120$^\circ$ geometry. These are Physik Instrumente LS-270 and LS-180 linear actuators, customised for reduced travel range and maximized stiffness\cite{RedaelliAO4ELT8}. The patrol range covers a $600\times300$~mm field, enabling acquisition of any NGS within the 180 arcsec MORFEO technical patrol field. The LO WFS arm operates in two configurations: a $2\times2$ Shack-Hartmann for accurate focus measurement in highly obscured fields and a full-aperture mode for optimal tip-tilt sensitivity and island-effect petal sensing \cite{PlantetAO4ELT8, 2022SPIE12185E..56A} thanks to the LIFT implementation. The final MORFEO baseline uses two FA/LIFT units and one $2\times2$ SH unit. The REF WFS is always an $8\times8$ SH in the visible. Table~\ref{tab:lor_params} summarises the key optical design parameters of the three WFS configurations as derived from the system optical model.

\begin{table}[ht]
\caption{\label{tab:lor_params} Key optical parameters of the LOR WFS configurations. F\# for the FA/LIFT LO mode refers to the image-plane F-number; for REF and $2\times2$ LO WFS it refers to the lenslet F-number. Throughput includes all optical surfaces and detector QE.}
\begin{center}
\setlength{\tabcolsep}{4pt}
\begin{tabular}{|l|c|c|c|}
\hline
\rule[-1ex]{0pt}{3.5ex} \textbf{Parameter} & \textbf{LO 1$\times$1 FA} & \textbf{LO 2$\times$2 SH} & \textbf{REF SH} \\
\hline
\rule[-1ex]{0pt}{3.5ex} Wavelength band    & $1.1\div1.8\mu m$ & $1.1\div1.8\mu m$ & $0.6\div1.0\mu m$  \\
\rule[-1ex]{0pt}{3.5ex} Subap. on diameter                         & 1        & 2         & 8         \\
\rule[-1ex]{0pt}{3.5ex} Pixels/subap.\ - after field stop          & 256      & 120       & 28        \\
\rule[-1ex]{0pt}{3.5ex} Pixel scale [mas\,pix$^{-1}$]              & 7        & 15        & 165       \\
\rule[-1ex]{0pt}{3.5ex} FoV [$''$] - after field stop                          & 1.79     & 1.80      & 4.62      \\
\rule[-1ex]{0pt}{3.5ex} F\# (image/lenslet)                        & 18.3     & 16.1      & 6.2       \\
\rule[-1ex]{0pt}{3.5ex} Throughput                                 & 32.5\%   & 32.5\%    & 21.5\%    \\
\rule[-1ex]{0pt}{3.5ex} WFS camera                                 & FREDA    & FREDA     & ALICE     \\
\hline
\end{tabular}
\end{center}
\end{table}

\noindent\textbf{Structural and thermoelastic analysis.}
A comprehensive FEA campaign verified structural integrity and optical stability over all arm positions and operational loads \cite{RedaelliAO4ELT8}. Modal analysis confirmed a first structural eigenfrequency of 24~Hz ($>$20~Hz required). A superposition model reconstructed the optical path difference (OPD) for all 12 arm configurations from a minimal set of FEA runs; Zemax propagation confirmed WFE contributions from gravity within the allocated budget ($\lesssim$50~nm\,RMS). Earthquake and FREDA vibration loads were verified against material yield. Thermoelastic analysis showed that a uniform $\pm$3\,°C variation produces a maximum centroid drift below 0.5~mas\,RMS, compliant with the $<$2.5~mas\,RMS pointing requirement.

\noindent\textbf{MAIT plan.}
The LOR MAIT follows a two-site strategy. At INAF--Merate, the LOR WFS boards undergo contact-metrology alignment using the Coord~3 CMM: all optical components are shimmed to their as-built nominal positions, and laser-tracker reference nests are installed as downstream datums. The aligned boards are then shipped to INAF--Arcetri for optical verification and refinement using the LOR AO Simulator (LAS) test source, installation and characterisation of the ESO-supplied FREDA and ALICE cameras, and final integration of the three LOR WFS Units onto the LOR Support Structure. After a successful Acceptance Readiness Review (ARR) at Arcetri, the complete module is delivered to INAF--Bologna for instrument-level AIT within the full MORFEO assembly.

\section{ADAPTIVE OPTICS CONTROL STRATEGY}
\label{sec:ao_control}
The MCAO loop is based on a Pseudo-Open-Loop Control (POLC) architecture. This strategy relies on accurate pseudo-open-loop slope estimation and advanced tomographic reconstructors optimized for varying atmospheric profiles \cite{AgapitoAO4ELT8}.

\subsection{Real-Time Control Architecture}
The computational load is managed by a Real-Time Computer split into two main subsystems, the Hard-Real Time Core (HRTC) and the Soft-Real Time Computer (SRTC). Both subsystems run on ESO-standard HW and SW platform: Commercial-Off the Shelf (COTS) computer servers and the ESO DevEnv software environment.

The HRTC is based on the Herzberg Extensible Adaptive Real-time Toolkit (HEART) \cite{2022SPIE12189E..26S} SW package. In MORFEO, the HEART processing pipelines handles pixel streams from up to 12 wavefront sensors in real-time, generating commands for the three DMs at high speed and low latency, in addition to commanding the six laser launchers jitter mirrors to remove the local LGS tilt. Calculations are performed on eight high-performance AMD EPYC processors, distributed across four dual-socket servers. Communication with all WFSs and DMs, as well with the telescope control system, is implemented using dedicated 10Gbit/s Ethernet links.

The Soft Real-Time Computer (SRTC) is based on the ESO RTC Toolkit \cite{eso_rtctk_2022}. It acts as the primary interface between the HRTC and the supervisory control software, managing background optimization tasks, telemetry data collection, and state management. It also estimates a number of system parameters used by the Instrument Control System (ICS) as an input for closing auxiliary loops. The SRTC also makes use of three NVIDIA Blackwell GPUs, in order to speed up critical calculations like the main LGS control matrix update and the fast update rate (10Hz) of the LIFT estimation.

Both the RTC and Instrument Control System (ICS) architectures have undergone substantial design updates to ensure seamless integration with MICADO \cite{CostaAO4ELT8}. Special attention has also been devoted to real-time regulator optimization for the NGS control loop and tip-tilt compensation, which is critical for maximizing sky coverage \cite{FaloticoAO4ELT8}.

\section{NUMERICAL SIMULATIONS AND PERFORMANCE ANALYSIS}
\label{sec:performance}
Extensive numerical simulations are mandatory to validate the MORFEO design against its top-level requirements. Recently, the main end-to-end simulation environment has transitioned from the PASSATA framework \cite{2016SPIE.9909E..7EA} to the next-generation SPECULA framework \cite{specula2026}. Other tools utilized for performance and error budget estimation include specific sky coverage analysis tools \cite{2022JATIS...8b1509P} and the analytical code TipTop \cite{2020SPIE11448E..2TN}.

The specific requirements defined in the MORFEO Technical Specification cover different atmospheric conditions and are summarized in Tab. \ref{tab:requirements}. The simulated values demonstrate compliance against the nominal requirements.

\begin{table}[ht]
\caption{Summary of the technical specification requirements and atmospheric conditions\cite{2013aoel.confE..89S}. In parenthesis the estimated SR, FWHM and EE values obtained via simulation. }
\label{tab:requirements}
\begin{center}
\begin{tabular}{|c|c|c|c|c|c|c|c|}
\hline
\rule[-1ex]{0pt}{3.5ex} \textbf{Case ID} & \textbf{Atmosp.} & \textbf{$\lambda$ [nm]} & \textbf{SR}  & \rule[-1ex]{0pt}{3.5ex} \textbf{$\Gamma$ [mas]}  & \textbf{EE [\%]} & \textbf{$\phi$ ["]} & \textbf{NGS asterism}\\
\hline
\rule[-1ex]{0pt}{3.5ex}  \multirow{3}{*}{R-MAO-80} & \multirow{3}{*}{Q1 (Best)} & 2200  & 0.60 (0.61) & 13 (12) & 38 (39) & \multirow{3}{*}{20} & \multirow{3}{*}{H=16.5, 20, b=40"} \\
\rule[-1ex]{0pt}{3.5ex} {} & {} & 1250 & 0.20 (0.23) & 9 (8) & 19 (24) & {} & {} \\
\rule[-1ex]{0pt}{3.5ex} {} & {} & 850 & 0.05 (0.05) & 8 (7) & 6 (11) & {} & {} \\
\hline
\rule[-1ex]{0pt}{3.5ex} \multirow{2}{*}{R-MAO-82} & \multirow{2}{*}{Median} & 2200 & 0.44 (0.45) & 13.5 (12.7) & 28 (30) & \multirow{2}{*}{60} & \multirow{2}{*}{H=17.5, 20, b=40"} \\
\rule[-1ex]{0pt}{3.5ex} {} & {} & 1250 & 0.08 (0.10) & 9 (9) & 10 (13) & {}  & {} \\
\hline
\rule[-1ex]{0pt}{3.5ex} R-MAO-83 & Q4 (Poor) & 2200 & 0.25 (0.29) & 14 (13) & 16 (21) & 20 & H=17.5, 20, b=40" \\
\hline
\rule[-1ex]{0pt}{3.5ex} R-MAO-168 & Q1 (1 NGS) & 2200 & 0.50 (0.51) & 14 (13) & 34 (35) & 20 & H=19, b=65" \\
\hline
\multicolumn{8}{c}{\scriptsize Note: $\lambda$ is the wavelength, $\phi$ is the science FoV diameter and $\Gamma$ is the FWHM. The Ensquared Energy (EE) is computed in a }\\\multicolumn{8}{c}{\scriptsize $16 \times 16$ mas aperture. In the the NGS asterism column: H is the H band magnitude and b is the asterism barycenter distance. }\\
\end{tabular}
\end{center}
\end{table}

\begin{figure}[ht]
    \begin{center}
        \subfigure[K band SR as a function of off-axis distance for the  requirements R-MAO-80, 82, 83 and 168.
        \label{fig:perf_z30}]
        {\includegraphics[width=0.45\columnwidth]{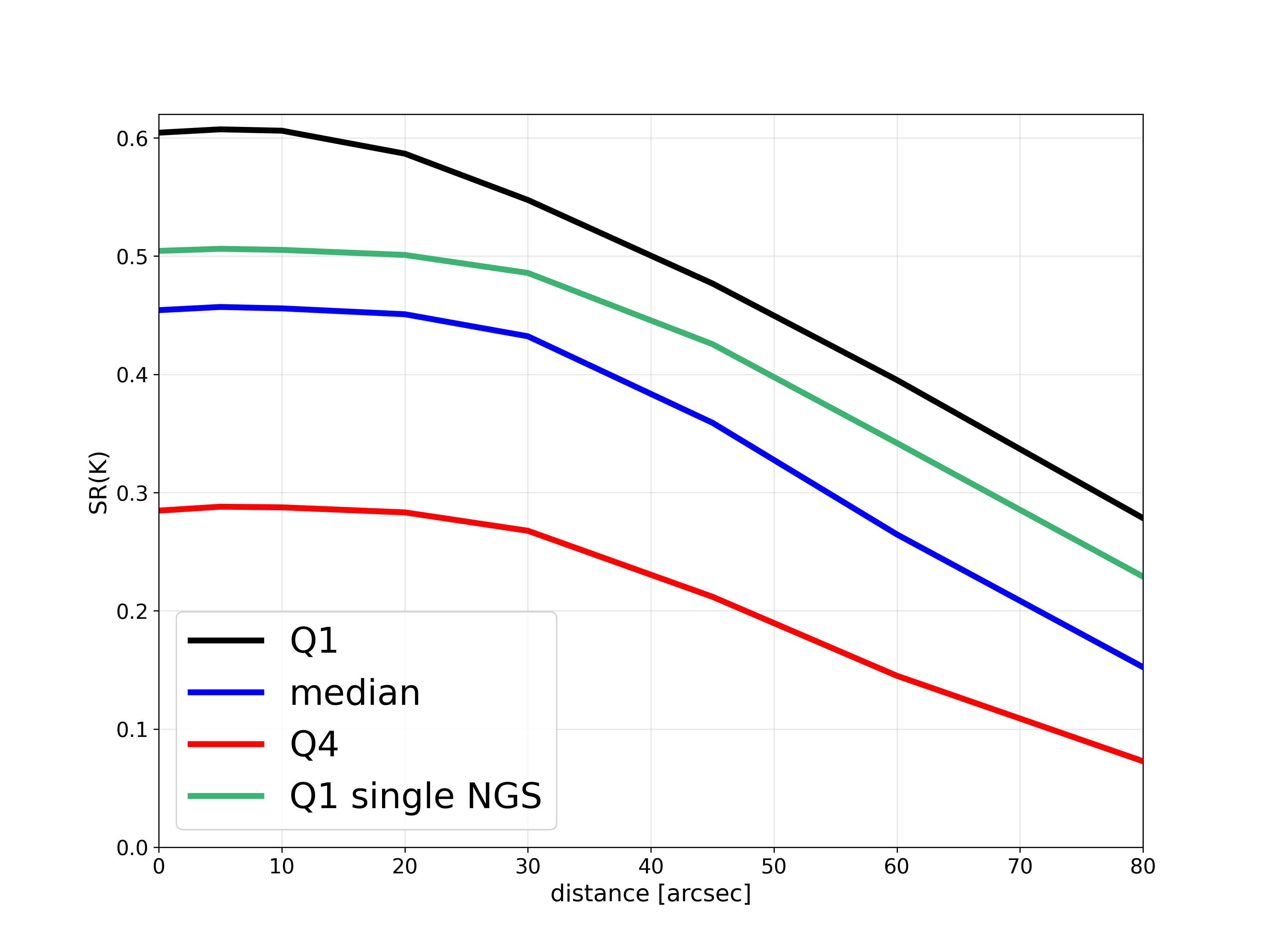}}
        \subfigure[On-axis K band SR as a function of the zenith angle for the median atmospheric profile.
        \label{fig:perf_z}]
        {\includegraphics[width=0.41\columnwidth]{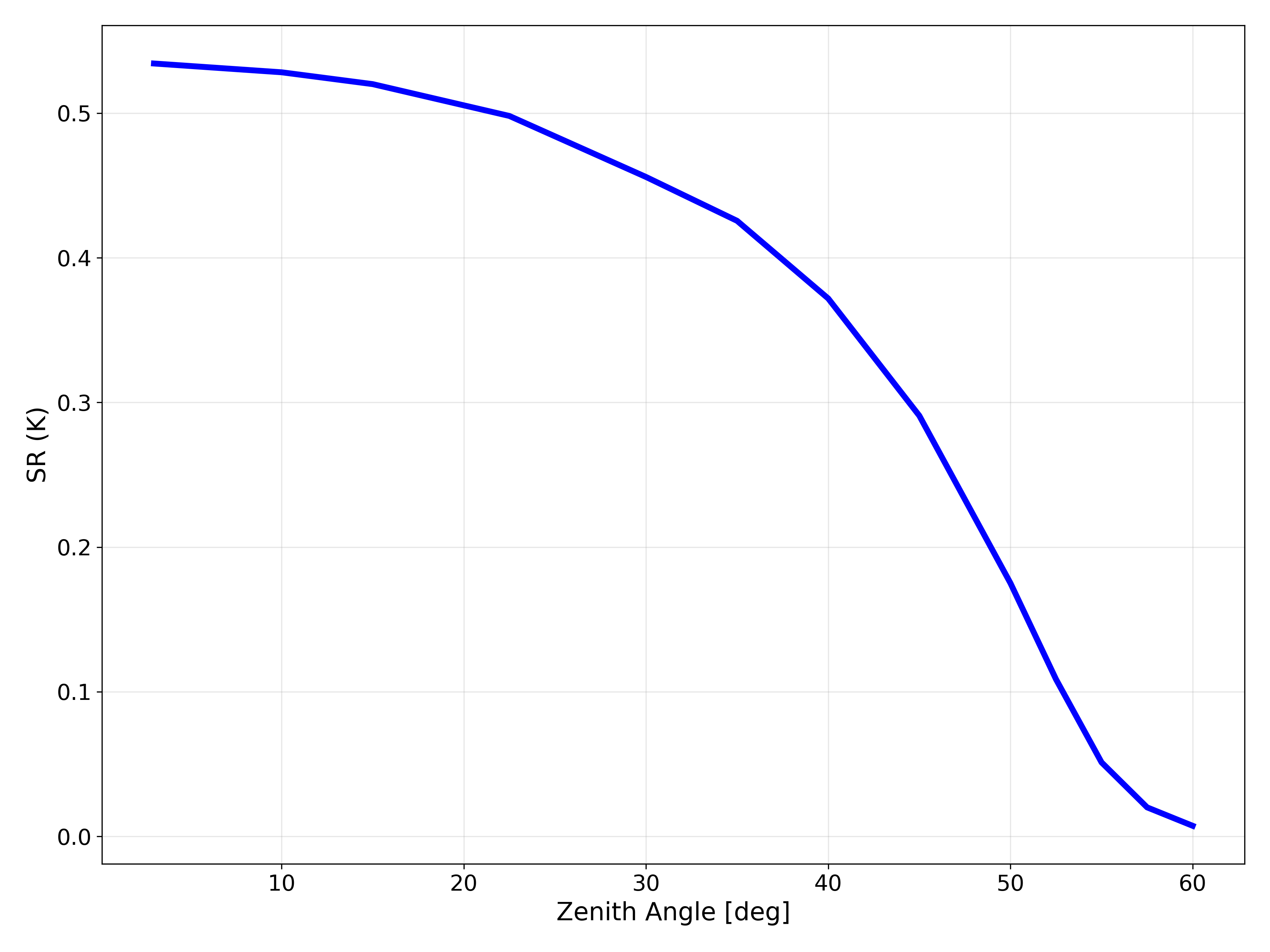}}
    \end{center}
   \caption{\label{fig:perf_req} MORFEO K band SR.}
\end{figure}

\begin{figure}[ht]
   \begin{center}
   \includegraphics[width=0.45\textwidth]{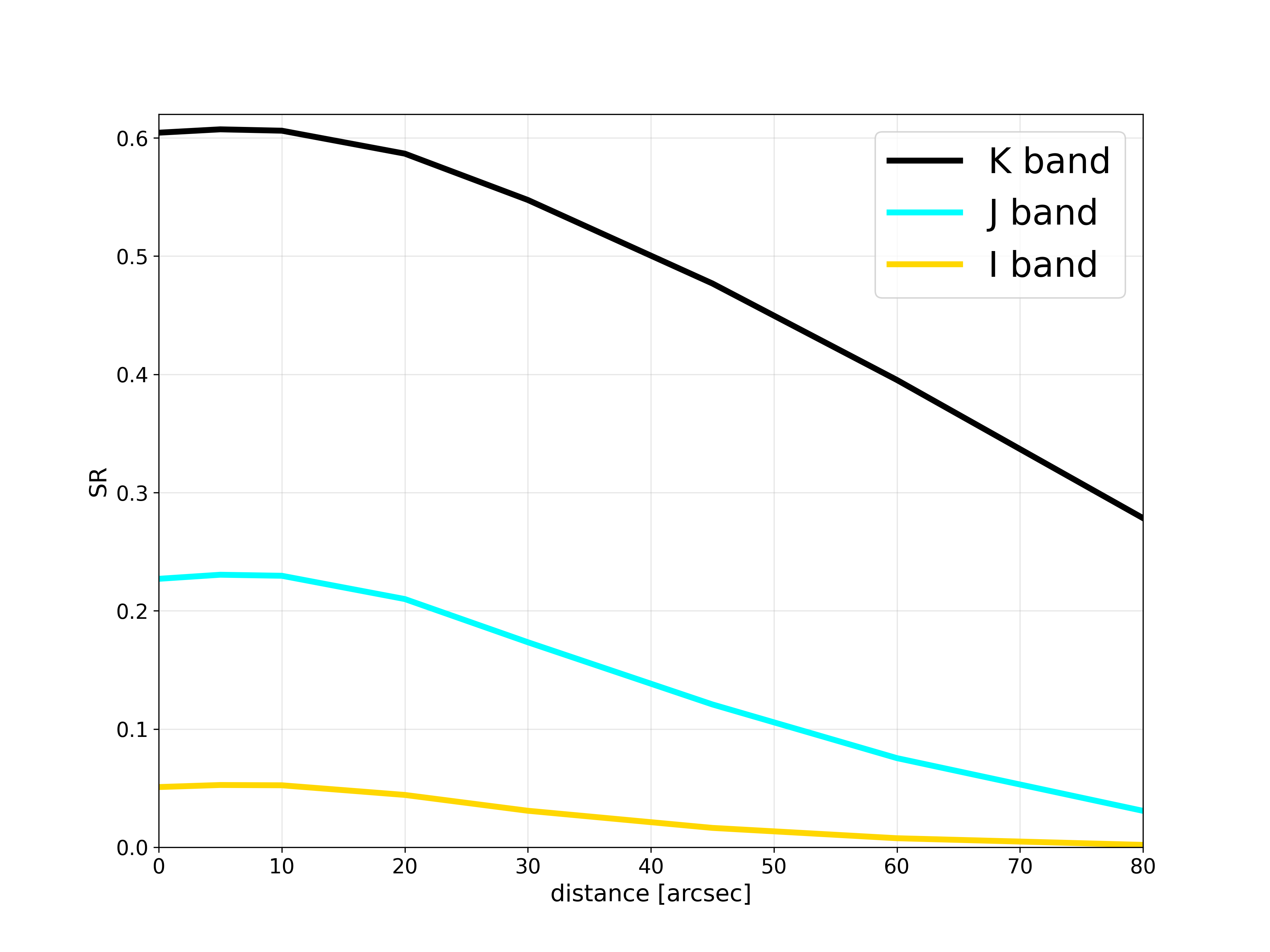}
   \end{center}
   \caption{\label{fig:SR_K_J_I} SR as a function of the off-axis distance for the Q1 atmospheric profile and I, J and K bands.}
\end{figure}

The estimation of system performance as a function of the zenith angle is presented in Fig.~\ref{fig:perf_req}, demonstrating the performance's dependence on the input seeing. The expected K, J, and I-band SRs for the Q1 atmospheric profile are further detailed in Fig.~\ref{fig:SR_K_J_I}.

\begin{figure}[!h]
    \begin{center}
        \subfigure[K band SR
        \label{fig:SC_SR}]
        {\includegraphics[width=0.32\columnwidth]{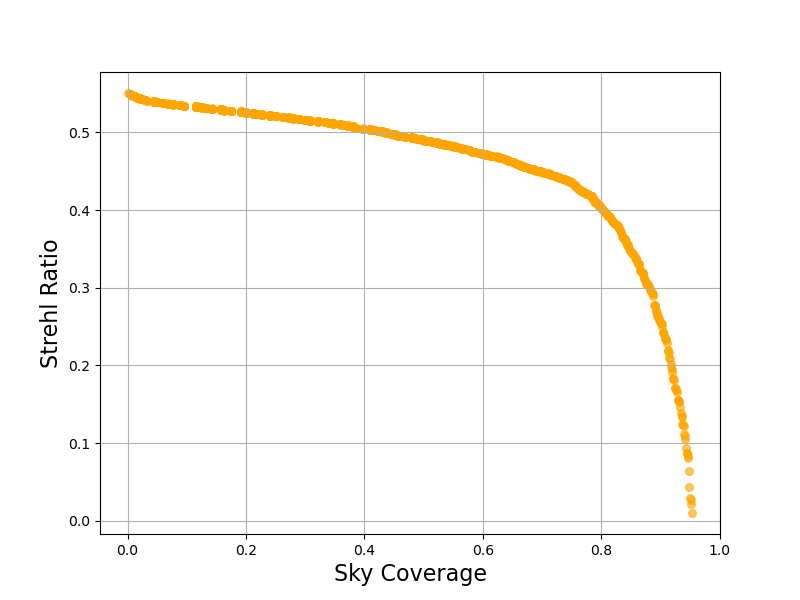}}
        \subfigure[K band FWHM
        \label{fig:SC_FWHM}]
        {\includegraphics[width=0.32\columnwidth]{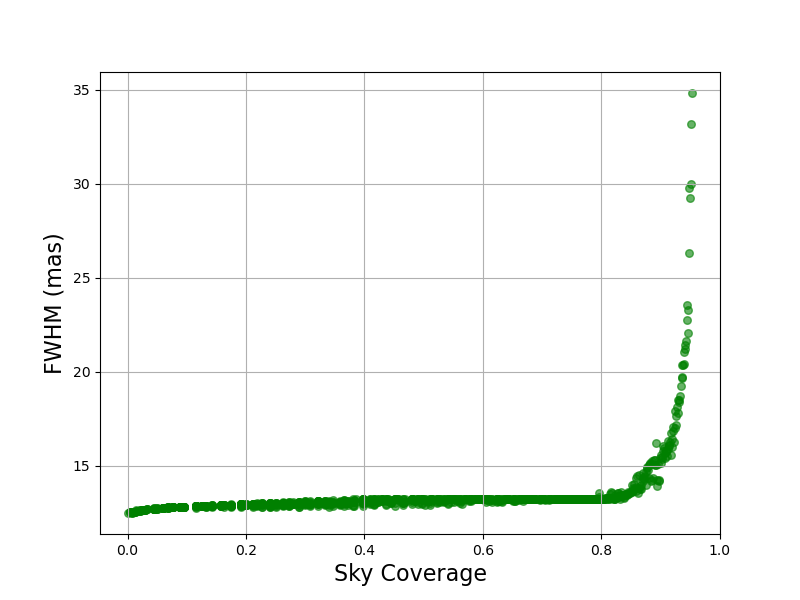}}
        \subfigure[K band EE (r=8mas)
        \label{fig:SC_EE}]
        {\includegraphics[width=0.32\columnwidth]{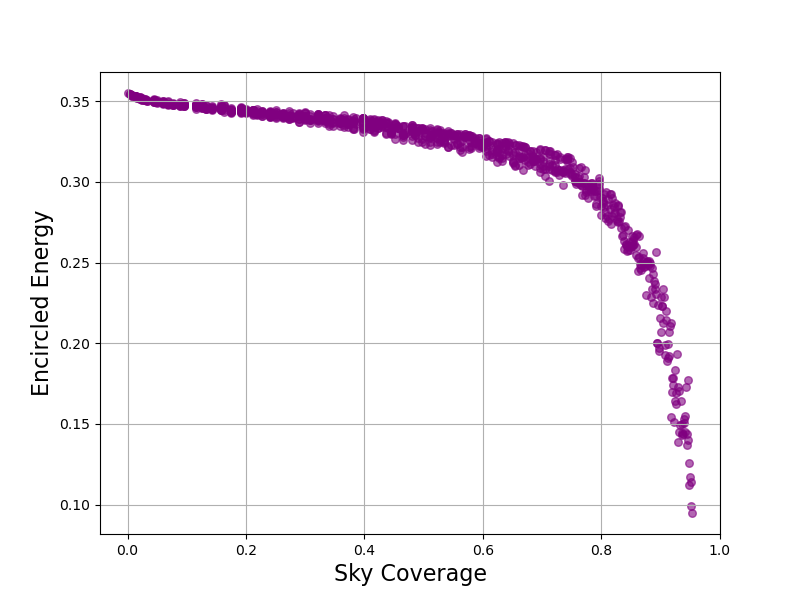}}
    \end{center}
    \caption{\label{fig:performance} Cumulative Distribution Function (the so-called \emph{sky coverage}) of the MORFEO performance in median atmospheric conditions (seeing 0.64 arcsec), at Z=30 deg over a r=30 arcsec scientific field considering 1000 pointings at South Galactic Pole.}
\end{figure}

\begin{figure}[!h]
    \begin{center}
        \subfigure[2D representation.
        \label{fig:drivers_2d}]
        {\includegraphics[width=0.45\columnwidth]{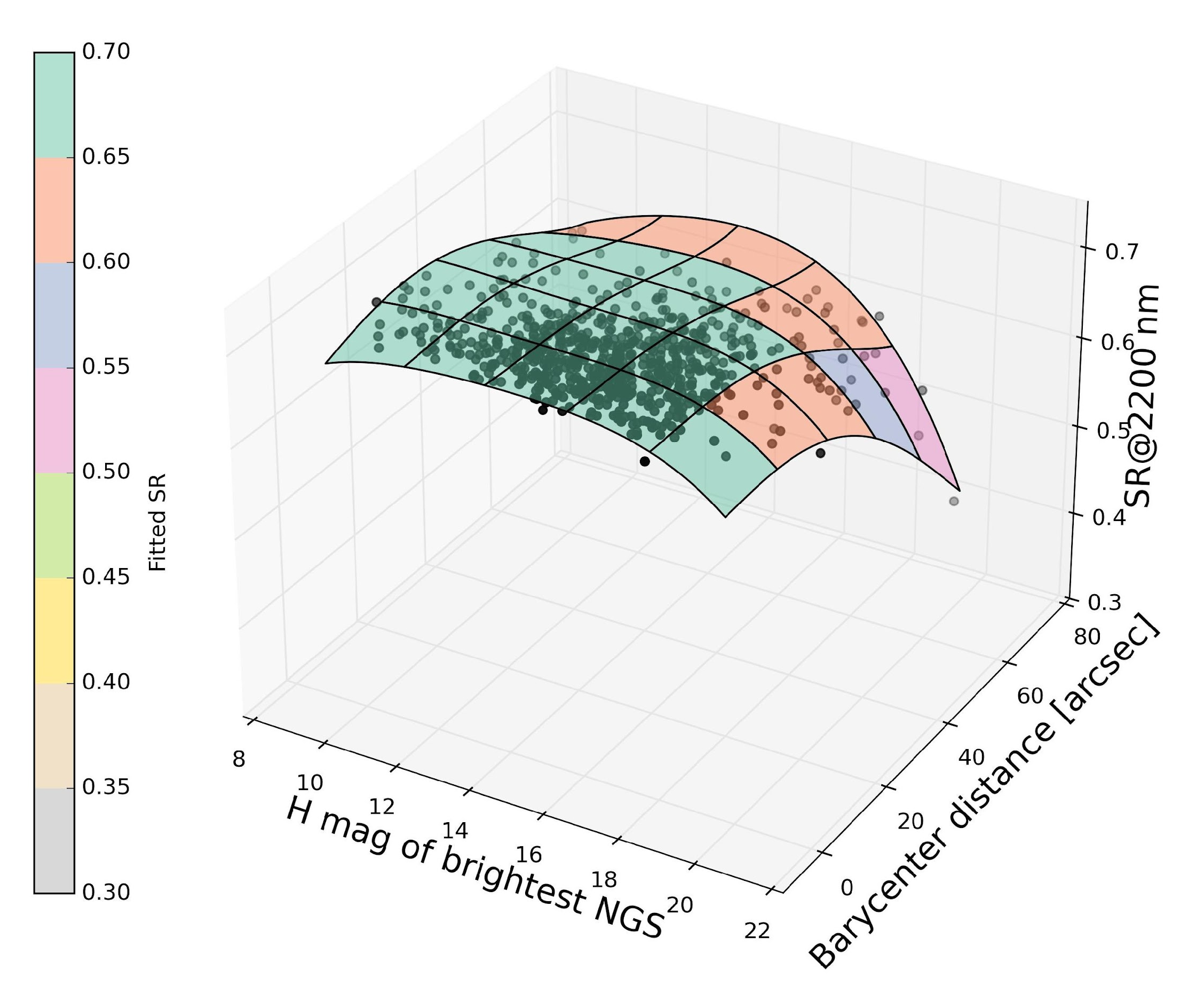}}
        \subfigure[1D representation.
        \label{fig:drivers_1d}]
        {\includegraphics[width=0.45\columnwidth]{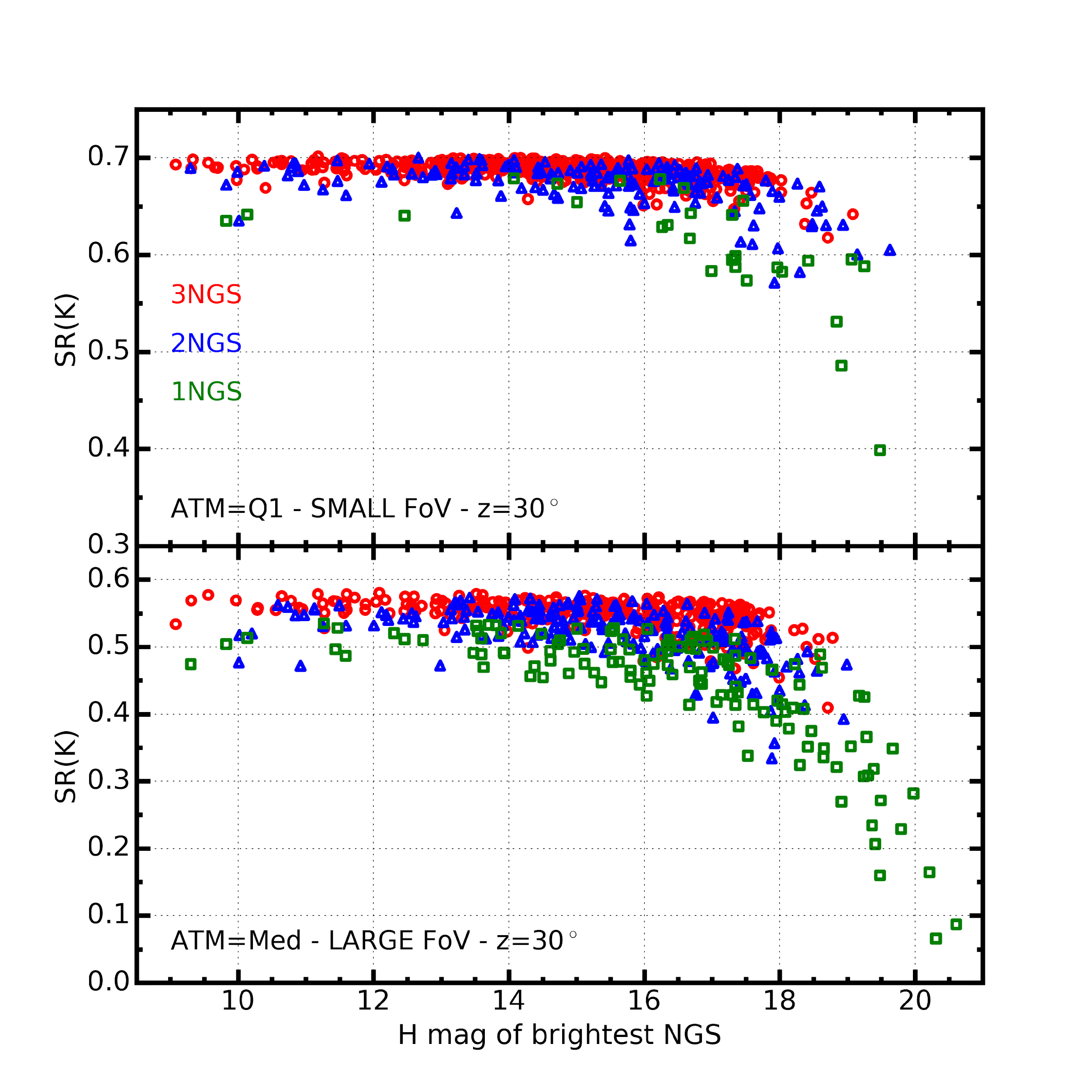}}
    \end{center}
    \caption{\label{fig:perf_driver} K band SR as a function of brightest NGS magnitude and NGS asterism barycenter distance for the Q1 atmospheric profile at Z=30 deg.}
\end{figure}

Fig.~\ref{fig:performance} provides a summary of the statistical performance under median atmospheric conditions. The system achieves stable performance over 80\% of the sky at the South Galactic Pole. While overall performance is heavily influenced by seeing and zenith angle, our sky coverage analysis (Fig.~\ref{fig:perf_driver}) indicates that when these atmospheric parameters are held constant, the magnitude of the brightest NGS is the dominant performance driver, followed closely by the geometry (barycenter distance) of the asterism. Interestingly, asterisms utilizing only two NGSs perform nearly identically to those utilizing three. 

A comprehensive breakdown of the updated overall wavefront error budget can be found in Agapito et al. 2026\cite{AgapitoAO4ELT8_2}.

\section{ROADMAP AND AIT PLAN}
\label{sec:roadmap}

To efficiently manage the project's high complexity, the Final Design Review (FDR) is structured as a multi-stage process rather than a single milestone. Several specific subsystem reviews have already been conducted or are currently closing their action items. These include the Optical FDRs (OFDR) for the main path and LGS, the Deformable Mirrors FDR, and the FDR1 for the RTC, LGS, and NGS modules. Concurrently, joint FDRs involving INAF, ESO, and industrial partners are addressing major hardware components, such as the Main Structure, the Calibration Unit, the LGSO, and the primary aspherical and flat mirrors.

The final comprehensive stage, FDR2, is scheduled to follow and will cover the remaining system aspects. This phase will encompass the Instrument Control System (both software and hardware), the remaining Adaptive Optics components, System Engineering, Science Operations, and Calibration. It will also systematically review specific hardware units—such as the Thermal Control System, the Test Unit, the dichroic, and the corrective plates—along with the overall Assembly, Integration, and Testing/Verification (AIT/V) procedures \cite{2024SPIE13097E..22C}.

As the design consolidates, the focus is shifting toward the Manufacturing, Assembly, Integration, and Testing (MAIT) phase. The AIV activities are planned to take place first in Europe, starting at the subsystems' premises and converging in the Bologna Integration Hall (BIH), a large facility recently refurbished to host the system when completely assembled. In BIH the system, equipped with the Test Unit (TU) to simulate the telescope and the Test and Alignment Camera (TAC) to simulate the instrument, will undergo the system tests. The Preliminary Acceptance in Europe (PAE) is scheduled for 2029, followed by shipping to Chile for further AIV activities in the Telescope Integration Hall and on the Nasmyth platform. These will be followed by the commissioning phases that will lead to the Provisional Acceptance in Chile (PAC) targeted for 2030, and Final Acceptance in Chile (FAC) in 2031 \cite{2024SPIE13097E..22C}.

\section{CONCLUSIONS}
\label{sec:conclusions}
The MORFEO project is on track toward completing its Final Design Review. The updated optical and mechanical designs, combined with robust AO control strategies and next-generation numerical simulations, confirm the system's capability to deliver the required wide-field, diffraction-limited correction for the ELT.

\printbibliography 

@ARTICLE{2021Msngr.182...13C,
       author = {{Ciliegi}, Paolo and {Agapito}, G. and {Aliverti}, M. and {Annibali}, F. and {Arcidiacono}, C. and {Balestra}, A. and {Baruffolo}, A. and {Bergomi}, M. and {Bianco}, A. and {Bonaglia}, M. and {Busoni}, L. and {Cantiello}, M. and {Cascone}, E. and {Chauvin}, G. and {Chinellato}, S. and {Cianniello}, V. and {Correia}, J. -J. and {Cosentino}, G. and {Dall'Ora}, M. and {De Caprio}, V. and {Devaney}, N. and {Di Antonio}, I. and {Di Cianno}, A. and {Di Giammatteo}, U. and {D'Orazi}, V. and {Di Rico}, G. and {Dolci}, M. and {Dout{\`e}}, S. and {Eredia}, C. and {Farinato}, J. and {Esposito}, S. and {Fantinel}, D. and {Feautrier}, P. and {Foppiani}, I. and {Giro}, E. and {Gluck}, L. and {Golden}, A. and {Goncharov}, A. and {Grani}, P. and {Gullieuszik}, M. and {Haguenauer}, P. and {H{\'e}nault}, F. and {Hubert}, Z. and {Le Louran}, M. and {Magrin}, D. and {Maiorano}, E. and {Mannucci}, F. and {Malone}, D. and {Marafatto}, L. and {Moraux}, E. and {Munari}, M. and {Oberti}, S. and {Pariani}, G. and {Pettazzi}, L. and {Plantet}, C. and {Podio}, L. and {Portaluri}, E. and {Puglisi}, A. and {Ragazzoni}, R. and {Rakich}, A. and {Rabou}, P. and {Redaelli}, E. and {Redman}, M. and {Riva}, M. and {Rochat}, S. and {Rodeghiero}, G. and {Salasnich}, B. and {Saracco}, P. and {Sordo}, R. and {Spavone}, M. and {Sztefek}, M. -H. and {Valentini}, A. and {Vanzella}, E. and {Verinaud}, C. and {Xompero}, M. and {Zaggia}, S.},
        title = "{MAORY: A Multi-conjugate Adaptive Optics RelaY for ELT}",
      journal = {The Messenger},
         year = 2021,
        month = mar,
       volume = {182},
        pages = {13-16},
          doi = {10.18727/0722-6691/5216},
archivePrefix = {arXiv},
       eprint = {2103.11219},
 primaryClass = {astro-ph.IM},
       adsurl = {https://ui.adsabs.harvard.edu/abs/2021Msngr.182...13C}
}

@ARTICLE{2021Msngr.182...17D,
       author = {{Davies}, R. and {H{\"o}rmann}, V. and {Rabien}, S. and {Sturm}, E. and {Alves}, J. and {Cl{\'e}net}, Y. and {Kotilainen}, J. and {Lang-Bardl}, F. and {Nicklas}, H. and {Pott}, J. -U. and {Tolstoy}, E. and {Vulcani}, B. and {MICADO Consortium}},
        title = "{MICADO: The Multi-Adaptive Optics Camera for Deep Observations}",
      journal = {The Messenger},
         year = 2021,
        month = mar,
       volume = {182},
        pages = {17-21},
          doi = {10.18727/0722-6691/5217},
archivePrefix = {arXiv},
       eprint = {2103.11631},
 primaryClass = {astro-ph.IM},
       adsurl = {https://ui.adsabs.harvard.edu/abs/2021Msngr.182...17D}
}

@ARTICLE{2024Msngr.192....3C,
       author = {{Cirasuolo}, M. and {Tamai}, R. and {Koehler}, B. and {Biancat-Marchet}, F. and {Gonz{\'a}les Herrera}, J. -C. and {ELT Team}},
        title = "{The Rise of the Giant: ESO's Extremely Large Telescope}",
      journal = {The Messenger},
         year = 2024,
        month = mar,
       volume = {192},
        pages = {3-3},
          doi = {10.18727/0722-6691/5346},
       adsurl = {https://ui.adsabs.harvard.edu/abs/2024Msngr.192....3C}
}

@INPROCEEDINGS{2024SPIE13097E..22C,
       author = {{Ciliegi}, Paolo and {Agapito}, Guido and {Aliverti}, Matteo and {Annibali}, Francesca and {Aridiacono}, Carmelo and {Azzaroli}, Nicol{\`o} and {Balestra}, Andrea and {Baronchelli}, Ivano and {Ballone}, Alessandro and {Baruffolo}, Andrea and {Battaini}, Federico and {Benedetti}, Simone and {Bergomi}, Maria and {Bianco}, Andrea and {Bonaglia}, Marco and {Briguglio}, Runa and {Busoni}, Lorenzo and {Cantiello}, Michele and {Capasso}, Giulio and {Carl{\`a}}, Giulia and {Carolo}, Elena and {Cascone}, Enrico and {Chauvin}, Ga{\"e}l. and {Chebbo}, Manal and {Chinellato}, Simonetta and {Cianniello}, Vincenzo and {Colapietro}, Mirko and {Correia}, Jean-Jacques and {Cosentino}, Giuseppe and {Costa}, Elia and {D'Auria}, Domenico and {De Caprio}, Vincenzo and {Devaney}, Nicholas and {Di Antonio}, Ivan and {Di Cianno}, Amico and {Di Dato}, Andrea and {Di Filippo}, Simone and {Di Francesco}, Benedetta and {Di Giammatteo}, Ugo and {Di Prospero}, Chiara and {Di Rico}, Gianluca and {Di Rocco}, Andrea and {Diretto}, Daphne and {Dolci}, Mauro and {Eredia}, Christian and {Esposito}, Simone and {Fantinel}, Daniela and {Farinato}, Jacopo and {Feautrier}, Philippe and {Foppiani}, Italo and {Genoni}, Matteo and {Giro}, Enrico and {Gluck}, Laurence and {Goncharov}, Alexander and {Grani}, Paolo and {Greggio}, Davide and {Guieu}, Sylvain and {Gullieuszik}, Marco and {Hubert}, Zoltan and {Jocou}, Laurent and {Lampitelli}, Salvatore and {Lapucci}, Tommaso and {Laudisio}, Fulvio and {Leal}, Vincent and {Magnard}, Yves and {Magrin}, Demetrio and {Malone}, Deborah and {Marafatto}, Luca and {Michel}, Christophe and {Mouillet}, David and {Moulin}, Thibaut and {Munari}, Matteo and {Oberti}, Sylvain and {Pancher}, Fabrice and {Pariani}, Giorgio and {Petrella}, Amedeo and {Pinnard}, Laurent and {Plantet}, Cedric and {Portaluri}, Elisa and {Puglisi}, Alfio and {Rabou}, Patrick and {Radhakrishnan}, Kalyan and {Ragazzoni}, Roberto and {Redaelli}, Edoardo Maria Alberto and {Riva}, Marco and {Rochat}, Sylvain and {Rodeghiero}, Gabriele and {Rosignoli}, Luca and {Salasnich}, Bernardo and {Savarese}, Salvatore and {Scalera}, Marcello and {Schipani}, Pietro and {Selvestrel}, Danilo and {Sassolas}, Benoit and {Sordo}, Rosanna and {Teodori}, Ludovico and {Umbriaco}, Gabriele and {Valentini}, Angelo and {Xompero}, Marco},
        title = "{MORFEO at ELT: the adaptive optics module for ELT}",
    booktitle = {Adaptive Optics Systems IX},
         year = 2024,
       editor = {{Jackson}, Kathryn J. and {Schmidt}, Dirk and {Vernet}, Elise},
       series = {Society of Photo-Optical Instrumentation Engineers (SPIE) Conference Series},
       volume = {13097},
        month = aug,
          eid = {1309722},
        pages = {1309722},
          doi = {10.1117/12.3019058},
       adsurl = {https://ui.adsabs.harvard.edu/abs/2024SPIE13097E..22C}
}

@ARTICLE{2022JATIS...8b1514F,
       author = {{Fusco}, Thierry and {Agapito}, Guido and {Neichel}, Benoit and {Oberti}, Sylvain and {Correia}, Carlos and {Haguenauer}, Pierre and {Plantet}, C{\'e}dric and {Pedreros}, Felipe and {Ke}, Zibo and {Costille}, Anne and {Jouve}, Pierre and {Busoni}, Lorenzo and {Esposito}, Simone},
        title = "{Key wavefront sensors features for laser-assisted tomographic adaptive optics systems on the Extremely Large Telescope}",
      journal = {Journal of Astronomical Telescopes, Instruments, and Systems},
         year = 2022,
        month = apr,
       volume = {8},
          eid = {021514},
        pages = {021514},
          doi = {10.1117/1.JATIS.8.2.021514},
archivePrefix = {arXiv},
       eprint = {2206.11383},
 primaryClass = {astro-ph.IM},
       adsurl = {https://ui.adsabs.harvard.edu/abs/2022JATIS...8b1514F}
}

@INPROCEEDINGS{2016SPIE.9909E..7EA,
       author = {{Agapito}, G. and {Puglisi}, A. and {Esposito}, S.},
        title = "{PASSATA: object oriented numerical simulation software for adaptive optics}",
    booktitle = {Adaptive Optics Systems V},
         year = 2016,
       editor = {{Marchetti}, Enrico and {Close}, Laird M. and {V{\'e}ran}, Jean-Pierre},
       series = {Society of Photo-Optical Instrumentation Engineers (SPIE) Conference Series},
       volume = {9909},
        month = jul,
          eid = {99097E},
        pages = {99097E},
          doi = {10.1117/12.2233963},
archivePrefix = {arXiv},
       eprint = {1607.07624},
 primaryClass = {astro-ph.IM},
       adsurl = {https://ui.adsabs.harvard.edu/abs/2016SPIE.9909E..7EA}
}

@INPROCEEDINGS{2022SPIE12185E..56A,
       author = {{Agapito}, Guido and {Busoni}, Lorenzo and {Carl{\`a}}, Giulia and {Plantet}, C{\'e}dric and {Esposito}, Simone and {Ciliegi}, Paolo},
        title = "{MAORY/MORFEO and LIFT: can the low order wavefront sensors become phasing sensors?}",
    booktitle = {Adaptive Optics Systems VIII},
         year = 2022,
       editor = {{Schreiber}, Laura and {Schmidt}, Dirk and {Vernet}, Elise},
       series = {Society of Photo-Optical Instrumentation Engineers (SPIE) Conference Series},
       volume = {12185},
        month = aug,
          eid = {1218556},
        pages = {1218556},
          doi = {10.1117/12.2629352},
archivePrefix = {arXiv},
       eprint = {2208.02662},
 primaryClass = {astro-ph.IM},
       adsurl = {https://ui.adsabs.harvard.edu/abs/2022SPIE12185E..56A}
}

@INPROCEEDINGS{2023aoel.confE..10A,
       author = {{Agapito}, Guido and {Busoni}, Lorenzo and {Plantet}, Cedric and {Carl{\`a}}, Giulia and {Bonaglia}, Marco and {Ciliegi}, Paolo},
        title = "{NGSs acquisition in MORFEO}",
    booktitle = {Adaptive Optics for Extremely Large Telescopes (AO4ELT7)},
         year = 2023,
        month = jun,
          eid = {10},
        pages = {10},
          doi = {10.13009/AO4ELT7-2023-011},
archivePrefix = {arXiv},
       eprint = {2310.08181},
 primaryClass = {astro-ph.IM},
       adsurl = {https://ui.adsabs.harvard.edu/abs/2023aoel.confE..10A}
}

@ARTICLE{2022JATIS...8b1505A,
       author = {{Agapito}, Guido and {Busoni}, Lorenzo and {Carl{\`a}}, Giulia and {Plantet}, C{\'e}dric and {Esposito}, Simone},
        title = "{Rolling shutter-induced aberrations in laser guide star wavefront sensing}",
      journal = {Journal of Astronomical Telescopes, Instruments, and Systems},
         year = 2022,
        month = apr,
       volume = {8},
          eid = {021505},
        pages = {021505},
          doi = {10.1117/1.JATIS.8.2.021505},
archivePrefix = {arXiv},
       eprint = {2208.02661},
 primaryClass = {astro-ph.IM},
       adsurl = {https://ui.adsabs.harvard.edu/abs/2022JATIS...8b1505A}
}

@INPROCEEDINGS{2020SPIE11448E..2TN,
       author = {{Neichel}, Benoit and {Beltramo-Martin}, Olivier and {Plantet}, C{\'e}dric and {Rossi}, Fabio and {Agapito}, Guido and {Fusco}, Thierry and {Carolo}, Elena and {Carl{\`a}}, Giulia and {Cirasuolo}, Michele and {Van Der Burg}, Remco},
        title = "{TIPTOP: a new tool to efficiently predict your favorite AO PSF}",
    booktitle = {Adaptive Optics Systems VII},
         year = 2020,
       editor = {{Schreiber}, Laura and {Schmidt}, Dirk and {Vernet}, Elise},
       series = {Society of Photo-Optical Instrumentation Engineers (SPIE) Conference Series},
       volume = {11448},
        month = dec,
          eid = {114482T},
        pages = {114482T},
          doi = {10.1117/12.2561533},
archivePrefix = {arXiv},
       eprint = {2101.06486},
 primaryClass = {astro-ph.IM},
       adsurl = {https://ui.adsabs.harvard.edu/abs/2020SPIE11448E..2TN}
}

@INPROCEEDINGS{2024SPIE13097E..4RP,
       author = {{Pariani}, G. and {Agapito}, G. and {Magrin}, D. and {Munari}, M. and {Busoni}, L. and {Riva}, M. and {Di Rocco}, A. and {Ciliegi}, P.},
        title = "{Interfacing adaptive optics simulations with the optical model: a powerful tool for MORFEO}",
    booktitle = {Adaptive Optics Systems IX},
         year = 2024,
       editor = {{Jackson}, Kathryn J. and {Schmidt}, Dirk and {Vernet}, Elise},
       series = {Society of Photo-Optical Instrumentation Engineers (SPIE) Conference Series},
       volume = {13097},
        month = aug,
          eid = {130974R},
        pages = {130974R},
          doi = {10.1117/12.3018817},
       adsurl = {https://ui.adsabs.harvard.edu/abs/2024SPIE13097E..4RP}
}

@ARTICLE{2022JATIS...8b1509P,
       author = {{Plantet}, C{\'e}dric and {Neichel}, Beno{\^\i}t and {Agapito}, Guido and {Busoni}, Lorenzo and {Correia}, Carlos M. and {Fusco}, Thierry and {Bonaglia}, Marco and {Esposito}, Simone},
        title = "{Sky coverage assessment for the European ELT: a joint evaluation for MAORY/MICADO and HARMONI}",
      journal = {Journal of Astronomical Telescopes, Instruments, and Systems},
         year = 2022,
        month = apr,
       volume = {8},
          eid = {021509},
        pages = {021509},
          doi = {10.1117/1.JATIS.8.2.021509},
       adsurl = {https://ui.adsabs.harvard.edu/abs/2022JATIS...8b1509P}
}

@INPROCEEDINGS{2013aoel.confE..89S,
       author = {{Sarazin}, Marc and {Le Louarn}, Miska and {Ascenso}, Joana and {Lombardi}, Giancluca and {Navarrete}, Julio},
        title = "{Defining reference turbulence profiles for E-ELT AO performance simulations}",
    booktitle = {Proceedings of the Third AO4ELT Conference},
         year = 2013,
       editor = {{Esposito}, Simone and {Fini}, Luca},
        month = dec,
          eid = {89},
        pages = {89},
          doi = {10.12839/AO4ELT3.13383},
       adsurl = {https://ui.adsabs.harvard.edu/abs/2013aoel.confE..89S}
}

@INPROCEEDINGS{2022SPIE12185E..4PA,
       author = {{Aliverti}, Matteo and {Pariani}, Giorgio and {Magrin}, Demetrio and {Redaelli}, Edoardo Maria Alberto and {Doniselli}, Simone and {Colapietro}, Mirko and {Salasnich}, Bernardo and {De Caprio}, Vincenzo and {Cianniello}, Vincenzo and {Eredia}, Christian and {Cascone}, Enrico and {Riva}, Marco and {Ciliegi}, Paolo and {Di Giammatteo}, Ugo},
        title = "{MAORY/MORFEO at ELT: Thermal Control System preliminary design}",
    booktitle = {Adaptive Optics Systems VIII},
         year = 2022,
       editor = {{Schreiber}, Laura and {Schmidt}, Dirk and {Vernet}, Elise},
       series = {Society of Photo-Optical Instrumentation Engineers (SPIE) Conference Series},
       volume = {12185},
        month = aug,
          eid = {121854P},
        pages = {121854P},
          doi = {10.1117/12.2629908},
       adsurl = {https://ui.adsabs.harvard.edu/abs/2022SPIE12185E..4PA}
}

@INPROCEEDINGS{2024SPIE13097E..54M,
       author = {{Magrin}, D. and {Pariani}, G. and {Munari}, M. and {Rabou}, P. and {Bianco}, A. and {Bergomi}, M. and {Redaelli}, E. and {Aliverti}, M. and {Riva}, M. and {Farinato}, J. and {Pinard}, L. and {Michel}, C. and {Sassolas}, B. and {Ballone}, A. and {Rodeghiero}, G. and {Carl{\`a}}, G. and {Greggio}, D. and {Busoni}, L. and {De Caprio}, V. and {Cianniello}, V. and {Kurita}, M. and {Teodori}, L. and {Di Rocco}, A. and {Ciliegi}, P.},
        title = "{Final optical design of MORFEO for the ELT}",
    booktitle = {Adaptive Optics Systems IX},
         year = 2024,
       editor = {{Jackson}, Kathryn J. and {Schmidt}, Dirk and {Vernet}, Elise},
       series = {Society of Photo-Optical Instrumentation Engineers (SPIE) Conference Series},
       volume = {13097},
        month = aug,
          eid = {1309754},
        pages = {1309754},
          doi = {10.1117/12.3020359},
       adsurl = {https://ui.adsabs.harvard.edu/abs/2024SPIE13097E..54M}
}

@INPROCEEDINGS{2024SPIE13096E..5LD,
       author = {{De Caprio}, Vincenzo and {Cianniello}, Vincenzo and {Eredia}, Christian and {D'Auria}, Domenico and {Cascone}, Enrico and {Redaelli}, Edoardo and {Aliverti}, Matteo and {Pariani}, Giorgio and {Riva}, Marco and {Farinato}, Jacopo and {Magrin}, Demetrio and {Marafatto}, Luca and {Chinellato}, Simonetta and {Rodeghiero}, Gabriele and {Di Rico}, Gianluca and {Teodori}, Ludovico and {Di Rocco}, Andrea and {Ciliegi}, Paolo},
        title = "{Main structure general overview: mechanical design update from PDR towards the final design phase}",
    booktitle = {Ground-based and Airborne Instrumentation for Astronomy X},
         year = 2024,
       editor = {{Bryant}, Julia J. and {Motohara}, Kentaro and {Vernet}, Jo{\"e}l. R.~D.},
       series = {Society of Photo-Optical Instrumentation Engineers (SPIE) Conference Series},
       volume = {13096},
        month = jul,
          eid = {130965L},
        pages = {130965L},
          doi = {10.1117/12.3019972},
       adsurl = {https://ui.adsabs.harvard.edu/abs/2024SPIE13096E..5LD}
}

@INPROCEEDINGS{2024SPIE13096E..5KM,
       author = {{Munari}, M. and {Magrin}, D. and {Pariani}, G. and {Rabou}, P. and {Farinato}, J. and {Ballone}, A. and {Bianco}, A. and {Aliverti}, M. and {Radaelli}, E. and {Riva}, M. and {Agapito}, G. and {Busoni}, L. and {De Caprio}, V. and {Cianniello}, V. and {Teodori}, L. and {Di Rocco}, A. and {Ciliegi}, P.},
        title = "{MORFEO LGSO optical design}",
    booktitle = {Ground-based and Airborne Instrumentation for Astronomy X},
         year = 2024,
       editor = {{Bryant}, Julia J. and {Motohara}, Kentaro and {Vernet}, Jo{\"e}l. R.~D.},
       series = {Society of Photo-Optical Instrumentation Engineers (SPIE) Conference Series},
       volume = {13096},
        month = jul,
          eid = {130965K},
        pages = {130965K},
          doi = {10.1117/12.3019893},
       adsurl = {https://ui.adsabs.harvard.edu/abs/2024SPIE13096E..5KM}
}

@article{specula2026,
  author = {Fabio Rossi and Alfio Puglisi and Guido Agapito},
  title = {{Introducing a new generation adaptive optics simulation framework: from PASSATA to SPECULA}},
  volume = {12},
  journal = {Journal of Astronomical Telescopes, Instruments, and Systems},
  number = {1},
  publisher = {SPIE},
  pages = {019001},
  year = {2026},
  doi = {10.1117/1.JATIS.12.1.019001},
  URL = {https://doi.org/10.1117/1.JATIS.12.1.019001}
}

@INPROCEEDINGS{2022SPIE12189E..26S,
       author = {{Smith}, Malcolm and {Chapin}, Ed and {Dunn}, Jennifer and {Gamroth}, Darryl and {Kerley}, Dan and {Mueller}, Lianne and {Stocks}, Jonathan and {V{\'e}ran}, Jean-Pierre},
        title = "{HEART: Herzberg Extensible Adaptive Real-time Toolkit (HEART): internal structure: blocks, pipes, and composition of a new RTC}",
    booktitle = {Software and Cyberinfrastructure for Astronomy VII},
         year = 2022,
       series = {Society of Photo-Optical Instrumentation Engineers (SPIE) Conference Series},
       volume = {12189},
        month = aug,
          eid = {1218926},
        pages = {1218926},
          doi = {10.1117/12.2630528},
       adsurl = {https://ui.adsabs.harvard.edu/abs/2022SPIE12189E..26S}
}

@INPROCEEDINGS{CostilleAO4ELT7,
       author = {{Costille}, Anne and {Renault}, Edgard and {Bonnefoi}, Anne and {Ceria}, William and {Dohlen}, Kjetil and {Hubert}, Zoltan and {Correia}, Jean-Jacques and {Moulin}, Thibaut and {Rabou}, Patrick and {Menendez Mendoza}, Saul and {Neichel}, Benoit and {Fusco}, Thierry and {El Hadi}, Kacem and {Curaba}, St{\'e}phane and {Clarke}, Fraser and {Melotte}, Dave and {Thatte}, Niranjan},
        title = "{Design and challenges for the HARMONI Laser Guide Star Sensors}",
    booktitle = {Adaptive Optics for Extremely Large Telescopes (AO4ELT7)},
         year = 2023,
        month = jun,
          eid = {9},
        pages = {9},
          doi = {10.13009/AO4ELT7-2023-010},
       adsurl = {https://ui.adsabs.harvard.edu/abs/2023aoel.confE...9C}
}

@INPROCEEDINGS{2022SPIE12185E..8DA,
       author = {{Agapito}, Guido and {Busoni}, Lorenzo and {Carl{\`a}}, Giulia and {Plantet}, C{\'e}dric and {Esposito}, Simone and {Ciliegi}, Paolo},
        title = "{MAORY/MORFEO and rolling shutter induced aberrations in laser guide star wavefront sensing}",
    booktitle = {Adaptive Optics Systems VIII},
         year = 2022,
       editor = {{Schreiber}, Laura and {Schmidt}, Dirk and {Vernet}, Elise},
       series = {Society of Photo-Optical Instrumentation Engineers (SPIE) Conference Series},
       volume = {12185},
        month = aug,
          eid = {121858D},
        pages = {121858D},
          doi = {10.1117/12.2629343},
archivePrefix = {arXiv},
       eprint = {2208.02661},
 primaryClass = {astro-ph.IM},
       adsurl = {https://ui.adsabs.harvard.edu/abs/2022SPIE12185E..8DA}
}

@INPROCEEDINGS{2018SPIE10703E..1YS,
       author = {{Schreiber}, Laura and {Feautrier}, Philippe and {Stadler}, Eric and {Rabou}, Patrick and {Correia}, Jean-Jacques and {Gl{\"u}ck}, Laurence and {Rochat}, Sylvain and {Jocou}, Laurent and {Magnard}, Yves and {Moulin}, Thibaut and {Delboulb{\'e}}, Alain and {Dout{\'e}}, Sylvain and {Chauvin}, Ga{\"e}l. and {Moraux}, Estelle and {Oberti}, Sylvain and {Verinaud}, Christophe and {Cortecchia}, Fausto and {Arcidiacono}, Carmelo and {Diolaiti}, Emiliano and {Ciliegi}, Paolo and {Bellazzini}, Michele and {Esposito}, Simone and {Busoni}, Lorenzo and {Ragazzoni}, Roberto},
        title = "{The MAORY laser guide star wavefront sensor: design status}",
    booktitle = {Adaptive Optics Systems VI},
         year = 2018,
       editor = {{Close}, Laird M. and {Schreiber}, Laura and {Schmidt}, Dirk},
       series = {Society of Photo-Optical Instrumentation Engineers (SPIE) Conference Series},
       volume = {10703},
        month = jul,
          eid = {107031Y},
        pages = {107031Y},
          doi = {10.1117/12.2314467},
       adsurl = {https://ui.adsabs.harvard.edu/abs/2018SPIE10703E..1YS}
}

@inproceedings{FaloticoAO4ELT8,
  title={{Real-Time Regulator Optimization for Tip-Tilt Compensation on Extremely Large Telescopes} },
  author={Falotico, Marco and others},
  Pages = {these proceedings}
}

@inproceedings{BusoniAO4ELT8,
  title={{MORFEO: Advancing Towards Final Design} },
  author={Busoni, Lorenzo and others},
  Pages = {these proceedings}
}

@inproceedings{PlantetAO4ELT8,
  title={{MORFEO and LIFT: are you allergic to petals?} },
  author={Plantet, C{\'e}dric and others},
  Pages = {these proceedings}
}

@unpublished{AgapitoAO4ELT8,
  title  = {{MORFEO control strategy}},
  author = {Agapito, Guido and others},
  note   = {in preparation}
}

@inproceedings{AgapitoAO4ELT8_2,
  title  = {{MORFEO wavefront error budget}},
  author = {Agapito, Guido and others},
  note   = {these proceedings}
}

@inproceedings{CostaAO4ELT8,
  title={{Advances in the Instrument Control System Software Design for MORFEO: Integration with MICADO and the post-focal Deformable Mirrors in the ELT Environment.} },
  author={Costa, Elia and others},
  Pages = {these proceedings}
}

@inproceedings{BiasiAO4ELT8,
  title={{Fifth generation AdOptica deformable mirrors for adaptive optics systems: Subaru ASM, GMT
ASMS, MORFEO DM1/DM2, VLT 2GDSM.} },
  author={Biasi, Roberto and others},
  Pages = {these proceedings}
}

@inproceedings{RedaelliAO4ELT8,
  title={{LOR optomechanical design and analyses.} },
  author={Redaelli, Edoardo and others},
  Pages = {these proceedings}
}

@misc{eso_rtctk_2022,
    author       = {{ESO}},
    title        = {{RTC Toolkit v.3.0.0 Documentation}},
    year         = {2022},
    url          = {https://ftp.eso.org/pub/elt/repos/docs/RTCTK/documents/v3.0.0/sphinx_doc/html/index.html} %,
    %note         = {\url{https://ftp.eso.org/pub/elt/repos/docs/RTCTK/documents/v3.0.0/sphinx_doc/html/index.html}}
}
\end{document}